\documentstyle[prl,aps]{revtex}

\begin{document}

\input epsf

\draft

\title{A doubled discretisation of abelian Chern--Simons theory}

\author{David H. Adams}

\address{School of Mathematics, Trinity College, Dublin 2, Ireland. E-mail:
dadams@maths.tcd.ie}

\maketitle

\begin{abstract}

A new discretisation of a doubled, i.e. BF, version of the pure abelian 
Chern--Simons theory is presented. It reproduces the continuum expressions
for the topological quantities of interest in the theory, namely the
partition function and correlation function of Wilson loops.
Similarities with free spinor field theory are discussed which are of interest
in connection with lattice fermion doubling.

\end{abstract}

\pacs{11.15.Ha, 2.40.Sf, 11.15.Tk}


The abelian Chern--Simons (CS) theory \cite{Schwarz,Jackiw,W(Jones)}
is an important topological field theory in three dimensions.
It provides the topological structure of topologically massive (abelian)
gauge theory \cite{Jackiw} and, in the euclidean metrics, provides a useful
theoretical framework for the description of interesting phenomena in planar
condensed matter physics as, for example, fractional statistics particles
\cite{anyons}, the quantum Hall effect, and high $T_c$ 
superconductivity \cite{Poly}. It is also
essentially the same as the weak coupling (large $k$) limit of the non-abelian
CS gauge theory, a solvable yet highly non-trivial topological
quantum field theory \cite{W(Jones)}. 
In this paper we describe a discretisation of the abelian CS theory
which reproduces the topological quantities of interest after introducing
a field doubling in the theory. This doubling leads to the abelian 
Chern--Simons action being replaced by the action for the 
so-called abelian BF gauge theory ((\ref{5}) below), in which the correlation
function of Wilson loops
and partition function become the square and norm-square respectively
of what they originally were.
Note that discretising the theory is not the same as putting it on a 
lattice in the usual way. Instead, it involves using a lattice to construct
a discrete {\em analogue} of the theory which reproduces the key topological
quantities and/or features, without having to take a continuum limit.
A detailed version of this work \cite{me} will be published elsewhere.

We take the spacetime to be euclidean ${\bf R}^3$
(the case of general 3-manifolds is dealt with in \cite{me}).
The abelian CS action for gauge field $A\!=\!A_{\mu}dx^{\mu}$ can 
be written as
\begin{eqnarray}
S(A)=\lambda\int_{\bf R^3}
dx\,\epsilon_{\mu\nu\rho}A_{\mu}\partial_{\nu}A_{\rho}
=\lambda\int_{\bf R^3}A\wedge{}dA=\lambda\langle{}A\,,(\ast{}d)A\rangle 
\label{4} 
\end{eqnarray}
where $\lambda$ is the coupling parameter,
$d$ is the exterior derivative, $\ast$ is the Hodge star operator, 
and $\langle\cdot\,,\cdot\rangle$ is the inner 
product in the space of 1-forms determined by euclidean metric in ${\bf R}^3$.
All the ingredients the last expression in (\ref{4}) have natural lattice
analogues (as we will see explicitly below); however the lattice analogue of
the operator $\ast$ is the {\em duality} operator, which maps between cells 
of the lattice $K$ and cells of the {\em dual} lattice $\widehat{K}$.
To accommodate this feature we introduce a new gauge field $A'$ 
and consider a doubled version of the action (\ref{4}):
\begin{eqnarray}
\widetilde{S}(A,A')\,\equiv\,
\lambda\langle\left({A \atop A'}\right)\,,\,
\left({0 \atop \ast{}d} \; {\ast{}d \atop 0} \right)
\left({A \atop A'}\right)\rangle 
=2\lambda\int_{\bf R^3}A'\wedge{}dA 
\label{5}
\end{eqnarray}
This is the action of the so-called abelian BF gauge theory \cite{Blau}.
In this theory the correlation function of {\em framed}
Wilson loops can be considered: A framed loop is a closed ribbon 
which we denote by $(\gamma,\gamma')$ where $\gamma$ and $\gamma'$ are the
two boundary loops of the ribbon. 
The Wilson correlation function of oriented framed 
loops $(\gamma^{(1)},\gamma^{(1)}{}'),\dots,(\gamma^{(r)},\gamma^{(r)}{}')$ is
\begin{eqnarray}
\langle(\gamma^{(1)},\gamma^{(1)}{}'),
\dots,(\gamma^{(r)},\gamma^{(r)}{}')\rangle
\equiv\widetilde{Z}(\lambda)^{-1}
\int_{{\cal A}\times{\cal A}}{\cal D\/}A{\cal D\/}A'\,
\lbrack\prod_{l=1}^r\Bigl(\,e^{i\oint_{\gamma^{(l)}}A}\Bigr)
\Bigl(\,e^{i\oint_{\gamma^{(l)}{}'}A'}\Bigr)
\rbrack\,e^{i\widetilde{S}(A,A')}
\label{6}
\end{eqnarray}
This can be formally evaluated using standard techniques \cite{Blau} to obtain
\begin{eqnarray}
\langle(\gamma^{(1)},\gamma^{(1)}{}'),
\dots,(\gamma^{(r)},\gamma^{(r)}{}')\rangle
=\exp\Bigl({\textstyle \frac{-i}{2\lambda}}\Bigl(\,\sum_{l\ne{}m}
L(\gamma^{(l)},\gamma^{(m)})+\sum_{l=1}^rL(\gamma^{(l)},\gamma^{(l)}{}')
\Bigr)\Bigr)
\label{7}
\end{eqnarray}
where $L(\gamma,\gamma')$ denotes the Gauss linking number 
of $\gamma$ and $\gamma'$.
The partition function of this theory,
\begin{eqnarray}
\widetilde{Z}(\lambda)\equiv
\int_{{\cal A}\times{\cal A}}{\cal D\/}A{\cal D\/}A'\,
e^{i\widetilde{S}(A,A')}\,,
\label{z0}
\end{eqnarray}
is also a quantity of topological interest.
After compactifying the spacetime to $S^3$ and imposing the the covariant
gauge-fixing condition
\begin{eqnarray}
d^{\dagger}A=0\qquad,\qquad{}d^{\dagger}A'=0\ ,
\label{z1}
\end{eqnarray}
(where $d^{\dagger}$ is the adjoint of $d$) the partition function
can be formally evaluated as in \cite{Schwarz} (see also \cite{AdSe})
to obtain
\begin{eqnarray}
\widetilde{Z}(\lambda)=\det(\phi_0^{\dagger}\phi_0)^{-1}\det{}'
(d_0^{\dagger}d_0)\det{}'
\Bigl({\textstyle\frac{-i\lambda}{\pi}
\left({0 \atop \ast{}d} \; {\ast{}d \atop 0} \right)}\Bigr)^{-1/2}
\label{z2}
\end{eqnarray}
where we denote by $d_q$ the restriction of $d$ to the space $\Omega^q(S^3)$
of $q$-forms, and $\phi_0:{\bf R}\to\Omega^0(S^3)$ maps $r\!\in\!{\bf R}$
to the constant function equal to $r$. In this expression 
$\det{}'(d_0^{\dagger}d_0)$ is the Faddeev--Popov determinant corresponding to
(\ref{z1}) and $\det(\phi_0^{\dagger}\phi_0)^{-1}$ ($=V(S^3)^{-1}$)
is a ``ghosts for ghosts'' 
determinant which arises because constant gauge transformations act trivially
on the gauge fields. 
The determinants in (\ref{z2}) are regularised via zeta-regularisation
as in \cite{Schwarz,AdSe}.
Using Hodge duality and the techniques of \cite{Schwarz,AdSe} we can
rewrite (\ref{z2}) as 
\begin{eqnarray}
\widetilde{Z}(\lambda)=\Bigl({\textstyle \frac{\lambda}{\pi}}\Bigr)^{-1}
\tau_{RS}(S^3;d)
\label{z3}
\end{eqnarray}
(the general phase factor of \cite[eq.(6)]{AdSe} is trivial here since the
operator in (\ref{5}) has symmetric spectrum) where
\begin{eqnarray}
\tau_{RS}(S^3;d)=\det(\phi_0^{\dagger}\phi_0)^{-1/2}
\det(\phi_3^{\dagger}\phi_3)^{1/2}
\prod_{q=0}^2\det{}'(d_q^{\dagger}d_q)^{\frac{1}{2}(-1)^q}
\label{z4}
\end{eqnarray}
is the Ray--Singer torsion of $d$ \cite{RS} (see in particular \S3
of the second paper in \cite{RS}).
(In (\ref{z4}) $\phi_3:{\bf R}\to\Omega^3(S^3)$ maps $r\!\in\!{\bf R}$ to the
harmonic 3-form $\omega$ with $\int_{S^3}\omega\!=\!r$ and
we have used $\det(\phi_3^{\dagger}\phi_3)\!=\!V(S^3)^{-1}\!=\!
\det(\phi_0^{\dagger}\phi_0)^{-1}$ \cite{me}.) The Ray--Singer torsion is a 
topological invariant of $S^3\,$, i.e. it is independent of the metric on 
$S^3$ used to construct $\ast$ and $\langle\cdot\,,\cdot\rangle$ in (\ref{5}),
$d^{\dagger}$ in (\ref{z1}), and $\phi_0^{\dagger}$ in (\ref{z2}). 
The physical significance of this is as follows:
When compactifying ${\bf R}^3$ to $S^3\approx{\bf R}^3\cup\{\infty\}$
(e.g. via stereographic projection) the euclidean metric on ${\bf R}^3$
must be deformed towards infinity in order that it extend to a well-defined
metric on $S^3$. The topological invariance of $\tau_{RS}(S^3,d)$ means
that the resulting partition function (\ref{z3})
is independent of how this deformation is carried out.
In fact $\tau_{RS}(S^3,d)\!=\!1$ (the argument for this will be given below)
so $\widetilde{Z}(\lambda)\!=\!\pi/\lambda$. 
If ${\bf R}^3$ is compactified in a topologically more complicated way,
leading to a general closed oriented 3-manifold $M\,$, then the preceding
derivation of (\ref{z3}) continues to hold (with $S^3$ replaced by $M$)
if $H^1(M)\!=\!0$ and can be generalised if $H^1(M)\ne0$ \cite{me}.
For example, if $M$ is a lens space $L(p,q)$ then $\tau_{RS}(L(p,q),d)\!=\!
1/p$ and $\widetilde{Z}(\lambda)\!=\!\pi/p\lambda$.

\begin{figure}
$$
\epsfysize=4cm \epsfbox{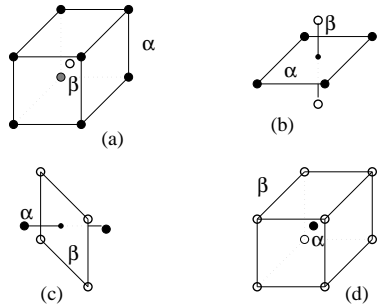}
$$
\caption{Definition of the duality operator $\ast^K\;$: $\alpha$ is a 
$p$-cell in $K$ and $\beta\!=\!\ast^K\alpha$ is the dual $(3\!-\!p)$-cell
in $\widehat{K}$. (In (a) $\beta$ is the point (vertex in $\widehat{K}$) 
at the centre of the 3-cell $\alpha\,$, while in (d) $\beta$ is the 3-cell in
$\widehat{K}$ which has the point $\alpha$ at its centre.)} 
\label{prl1}
\end{figure}

We will construct a discrete version of the doubled theory 
$\widetilde{S}(A,A')$ which reproduces the continuum expressions (\ref{7})
and (\ref{z3}) for the correlation function of framed Wilson loops and 
partition function respectively. Let $K$ be a lattice decomposition of
${\bf R}^3$ which, for convenience, we take to be cubic. It is well-known
\cite{Joos,discreteCS} that the space $\Omega^p$ of antisymmetric tensor 
fields of degree $p$ (i.e. $p$-forms) has a discrete analogue,
the space $C^p(K)$ of $p$-cochains (i.e. ${\bf R}$-valued functions on the 
$p$-cells of $K$), in particular $C^1(K)$ is the analogue of the space
${\cal A}\!=\!\Omega^1$ of gauge fields. The space $C_p(K)$ of $p$-chains
(i.e. formal linear combinations over ${\bf R}$ of oriented $p$-cells)
has a canonical inner product $\langle\cdot\,,\cdot\rangle$ defined by
requiring that the $p$-cells be orthonormal; this allows to identify
$C_p(K)$ with its dual space $C^p(K)$ so we will speak only of $C_p(K)$
in the following. The analogue of $d$
is the coboundary operator $d^K:C_p(K)\to{}C_{p+1}(K)\,$, i.e. the adjoint
of the boundary operator $\partial^K$. 
The new feature of our discretisation is that we also use the (co)chain
spaces $C_q(\widehat{K})$ associated with the {\em dual} lattice $\widehat{K}$
(i.e. the cubic lattice whose vertices are the centres of the 3-cells of $K$).
The cells of $K$ and $\widehat{K}$ are related by the {\em duality operator}
$\ast^K\,$, defined in Fig.\ \ref{prl1}.
An orientation for a $p$-cell $\alpha$ determines 
an orientation for the dual $(3\!-\!p)$-cell $\ast^K\alpha$ by requiring
that the product of the orientations of $\alpha$ and $\ast^K\alpha$ 
coincides with the standard orientation of ${\bf R}^3$. 
Thus the duality operator $\ast^K$ determines a linear map 
$\ast^K:C_p(K)\stackrel{\simeq}{\to}C_{3-p}(\widehat{K})\,$; this 
is the discrete analogue of the Hodge star operator $\ast$
in (\ref{4})--(\ref{5}).
Set $\ast^{\widehat{K}}\equiv(\ast^K)^{\dagger}\!=\!(\ast^K)^{-1}$.
The discrete theory is now constructed by
\begin{eqnarray}
(A,A')\in{\cal A}\times{\cal A}&\longrightarrow&(a,a')
\in{}C_1(K)\times{}C_1(\widehat{K}) \label{9a} \\
\widetilde{S}(A,A')\,\equiv\,
\lambda\Big\langle\left({A \atop A'}\right)\,,\,
\left({0 \atop \ast{}d} \; {\ast{}d \atop 0} \right)
\left({A \atop A'}\right)\Big\rangle 
&\longrightarrow&\widetilde{S}_K(a,a')\,\equiv\,
\lambda\Big\langle\left({a \atop a'}\right)\,,\,
\left( {0 \atop \ast^Kd^K} \; 
{\ast^{\widehat{K}}d^{\widehat{K}} \atop 0} \right)
\left({a \atop a'}\right)\Big\rangle
\label{10}
\end{eqnarray}
The discrete action $S_K(a,a')$ is invariant
under $a\to{}a\!+\!d^Kb\,$, $\;a'\to{}a'\!+\!d^{\widehat{K}}b'$ for all
$(b,b')\in{}C_0(K)\times{}C_0(\widehat{K})$ since $d^Kd^K\!=\!0$ and
$d^{\widehat{K}}d^{\widehat{K}}\!=\!0\,$; this is the discrete analogue
of the gauge invariance of the continuum theory.

\begin{figure}
$$
\epsfxsize=7cm \epsfbox{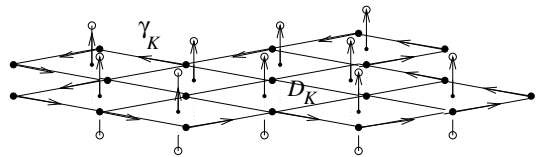}
$$
\caption{$\gamma_K$ is the boundary of the surface $D_K$ made up of 2-cells
of $K$. The vertical line segments are the duals of the 2-cells making
up $D_K$.}
\label{prl2}
\end{figure}

Framed Wilson loops fit naturally into this discrete setup: The framed loops
are taken to be ribbons $(\gamma_K,\gamma_{\widehat{K}})$ where one boundary
loop $\gamma_K$ is an edge loop in the lattice $K$ and the other boundary
loop $\gamma_{\widehat{K}}$ is an edge loop in the dual lattice $\widehat{K}$.
(It is always possible to 
find such a framing of an edge loop $\gamma_K$ \cite{me}.)
There is a natural discrete version of line integrals:
\begin{eqnarray}
\oint_{\gamma_K}A\ \longrightarrow\ \langle\underline{\gamma}_K\,,a\rangle
\qquad,\qquad \oint_{\gamma_{\widehat{K}}}A'\ \longrightarrow\ 
\langle\underline{\gamma}_{\widehat{K}}\,,a'\rangle 
\label{11}
\end{eqnarray}
where $\underline{\gamma}_K\in{}C_1(K)$ denotes the sum of the 1-cells
in $K$ making up $\gamma_K\,$, and $\underline{\gamma}_{\widehat{K}}\in{}
C_1(\widehat{K})$ denotes the sum of the 1-cells in $\widehat{K}$ making
up $\gamma_{\widehat{K}}$. 
Then the correlation function of non-intersecting oriented framed edge loops 
$(\gamma_K^{(1)},\gamma_{\widehat{K}}^{(1)}),\dots,
(\gamma_K^{(r)},\gamma_{\widehat{K}}^{(r)})$
in the discrete theory is
\begin{eqnarray}
\langle(\gamma_K^{(1)},\gamma_{\widehat{K}}^{(1)}),\dots,
(\gamma_K^{(r)},\gamma_{\widehat{K}}^{(r)})\rangle 
\equiv\widetilde{Z}_K(\lambda)^{-1}
\int_{C_1(K)\times{}C_1(\widehat{K})}{\cal D\/}a{\cal D\/}a'\,
\lbrack\prod_{l=1}^r\Bigl(\,e^{i\langle
\underline{\gamma}_K^{(l)}\,,a\rangle}\Bigr)
\Bigl(\,e^{i\langle\underline{\gamma}_{\widehat{K}}^{(l)}\,,a'\rangle}\Bigr)
\rbrack\,e^{i\widetilde{S}_K(a,a')}
\label{12}
\end{eqnarray}
A formal evaluation analogous to the 
evaluation of (\ref{6}) leading to (\ref{7}) gives
\begin{eqnarray}
\langle(\gamma_K^{(1)},\gamma_{\widehat{K}}^{(1)}),\dots,
(\gamma_K^{(r)},\gamma_{\widehat{K}}^{(r)})\rangle
&=&\exp\Bigl({\textstyle \frac{-i}{4\lambda}
\Big\langle\biggl({\underline{\gamma}_K^{(1)}+\dots+
\underline{\gamma}_K^{(r)} \atop
\underline{\gamma}_{\widehat{K}}^{(1)}+\dots+
\underline{\gamma}_{\widehat{K}}^{(r)} } \biggr)\,,\,
\left({0 \atop \ast^Kd^K} \; {\ast^{\widehat{K}}d^{\widehat{K}} \atop 0}
\right)^{\!-1}
\biggl({\underline{\gamma}_K^{(1)}+\dots+\underline{\gamma}_K^{(r)} \atop
\underline{\gamma}_{\widehat{K}}^{(1)}+\dots+
\underline{\gamma}_{\widehat{K}}^{(r)}}\biggr)\Big\rangle }\Bigr) \nonumber \\
&=&\exp\Bigl({\textstyle \frac{-i}{2\lambda}}\sum_{l,m=1}^r
\langle\underline{\gamma}_K^{(l)}\,,
(\ast^Kd^K)^{-1}\underline{\gamma}_{\widehat{K}}^{(m)}\rangle\Bigr)
\label{13}
\end{eqnarray}
where we have used $(\ast^Kd^K)^{\dagger}\!=\!
\ast^{\widehat{K}}d^{\widehat{K}}$.
To show that this coincides with the continuum expression (\ref{7})
we must show that for any oriented edge loop $\gamma_K$ in $K$ 
and oriented edge loop $\gamma_{\widehat{K}}$ in $\widehat{K}\,$,
\begin{eqnarray}
\langle\underline{\gamma}_K\,,(\ast^Kd^K)^{-1}
\underline{\gamma}_{\widehat{K}}\rangle
\;=\,L(\gamma_K,\gamma_{\widehat{K}})\,.
\label{14}
\end{eqnarray}
Then taking $\gamma_K\!=\!\gamma_K^{(l)}$ and $\gamma_{\widehat{K}}\!=\!
\gamma_{\widehat{K}}^{(m)}$ in (\ref{14}) and substituting in (\ref{13})
reproduces the continuum expression (\ref{7}).
To derive (\ref{14}) we
recall that the linking number of $\gamma_K$ and $\gamma_{\widehat{K}}$
can be characterised as follows. Let $D$
be a surface in ${\bf R}^3$ with $\gamma_K$ as its boundary, and such
that all intersections of $D$ with $\gamma_{\widehat{K}}$ are transverse, then 
\begin{eqnarray}
L(\gamma_K,\gamma_{\widehat{K}})=\sum_{D\,\cap\,\gamma_{\widehat{K}}}\pm1
\label{15}
\end{eqnarray}
where the sign of $\pm1$ for a given intersection of $D$ and 
$\gamma_{\widehat{K}}$ is $+$ if the product
of the orientations of $D$ (induced by the orientation of $\gamma_K$)
and $\gamma_{\widehat{K}}$ at the intersection coincides with the standard 
orientation of ${\bf R}^3\,$, and $-$ otherwise.
We now show that the l.h.s. of (\ref{14}) coincides with (\ref{15}).
First note that
\begin{eqnarray}
\langle\underline{\gamma}_K\,,
(\ast^Kd^K)^{-1}\underline{\gamma}_{\widehat{K}}\rangle
\;=\;\langle((\ast^Kd^K)^{-1})^{\dagger}\underline{\gamma}_K\,,
\underline{\gamma}_{\widehat{K}}\rangle
\;=\;\langle\ast^K(\partial^K)^{-1}\underline{\gamma}_K\,,
\underline{\gamma}_{\widehat{K}}\rangle\,.
\label{16}
\end{eqnarray}
Choose a surface $D_K$ in ${\bf R}^3$ made up of a union of 2-cells of $K$
and with $\gamma_K$ as its boundary (illustrated in Fig.\ \ref{prl2};
such a choice is always possible \cite{me})
and equip $D_K$ with the orientation induced by $\gamma_K$.
The formal sum of the oriented 2-cells making up $D_K$ is then an element
$\underline{D}_K\in{}C_2(K)\,$, and $\partial^K\underline{D}_K\!=\!
\underline{\gamma}_K\,$, so (\ref{16}) gives
\begin{eqnarray}
\langle\underline{\gamma}_K\,,
(\ast^Kd^K)^{-1}\underline{\gamma}_{\widehat{K}}\rangle
\;=\;\langle\ast^K\underline{D}_K\,,\underline{\gamma}_{\widehat{K}}\rangle
\label{17}
\end{eqnarray}
Now $\ast^K\underline{D}_K\in{}C_1(\widehat{K})$ is the sum of all the 1-cells
in $\widehat{K}$ which are dual to the 2-cells making up $D_K\,$, as 
indicated in Fig.\ \ref{prl2}.
Since $\gamma_{\widehat{K}}$ is an edge loop in the dual lattice $\widehat{K}$
all the 1-cells $\beta$ making up $\gamma_{\widehat{K}}$ are duals of
2-cells $\alpha$ in $K$ as illustrated in 
Fig.\ \ref{prl1}(b) above. Hence intersections
of $\gamma_{\widehat{K}}$ and $D_K$ occur precisely when a 1-cell in
$\gamma_{\widehat{K}}$ is the dual of a 2-cell in $D_K$ (up to a sign) and
it follows that the r.h.s. of (\ref{17}) equals (\ref{15}) with 
$D\!=\!D_K$. This completes the derivation of (\ref{14}), 
thereby showing that the Wilson
correlation function (\ref{13}) in the discrete theory reproduces the continuum
expression (\ref{7}) as claimed.

The partition function in this discrete theory is 
\begin{eqnarray}
\widetilde{Z}_K(\lambda)\equiv\int_{C_1(K)\times{}C_1(\widehat{K})}{\cal D\/}a
{\cal D\/}a'\,e^{i\widetilde{S}_K(a,a')}\,.
\label{w0}
\end{eqnarray}
As before we compactify the spacetime to $S^3\,$; taking $K$ to be a lattice
decomposition for $S^3$ the analogue of the gauge-fixing 
condition (\ref{z1}) is 
\begin{eqnarray}
\partial^Ka=0\qquad,\qquad\partial^{\widehat{K}}a'=0\ ,
\label{w1}
\end{eqnarray}
and a formal evaluation of (\ref{w0}) analogous to the one leading to 
(\ref{z2}) gives
\begin{eqnarray}
\widetilde{Z}(\lambda)=\det((\phi_0^K)^{\dagger}\phi_0^K)^{-1/2}\det((
\phi_0^{\widehat{K}})^{\dagger}\phi_0^{\widehat{K}})^{-1/2} 
\det{}'(\partial_1^Kd_0^K)^{1/2}\det{}'(\partial_1^{\widehat{K}}
d_0^{\widehat{K}})^{1/2}\det{}'\Bigl({\textstyle \frac{-i\lambda}{\pi}
\left({0 \atop \ast^Kd_1^K} \; {\ast^{\widehat{K}}d_1^{\widehat{K}} \atop 0} 
\right)}\Bigr)^{-1/2}
\label{w2}
\end{eqnarray}
Here $\phi_0^K:{\bf R}\to{}C_0(K)$ and $\phi_0^{\widehat{K}}:{\bf R}\to{}
C_0(\widehat{K})$ are natural discrete analogues of $\phi_0\,$; 
$\,\det((\phi_0^K)^{\dagger}\phi_0^K)\!=\!N_0^K$ and 
$\det((\phi_0^{\widehat{K}})^{\dagger}\phi_0^{\widehat{K}})\!=\!
N_0^{\widehat{K}}$ where 
$N_p^K\equiv\dim{}C_p(K)\,$, $\,N_q^{\widehat{K}}\!=\!\dim{}C_q(\widehat{K})$
\cite{me}. There is a natural discrete analogue $\phi_3^K$ of $\phi_3$ with 
$\det((\phi_3^K)^{\dagger}\phi_3^K)\!=\!1/N_3^K\!=\!1/N_0^{\widehat{K}}\!=\!
\det((\phi_0^{\widehat{K}})^{\dagger}\phi_0^{\widehat{K}})^{-1}\,$; 
using this and the properties of $\ast^K$ ($\partial_q^K\!=\!(-1)^q
\ast^{\widehat{K}}d_{3-q}^{\widehat{K}}\ast^K$) 
we can rewrite (\ref{w2}) as \cite{me}
\begin{eqnarray}
\widetilde{Z}_K(\lambda)=\Bigl({\textstyle \frac{\lambda}{\pi}}
\Bigr)^{-1+N_0^K-N_1^K}\tau(S^3;K,d^K)
\label{w3}
\end{eqnarray}
where 
\begin{eqnarray}
\tau(S^3;K,d^K)=\det((\phi_0^K)^{\dagger}\phi_0^K)^{-1/2}
\det((\phi_3^K)^{\dagger}\phi_3^K)^{1/2} 
\prod_{q=0}^2\det{}'(\partial_{q+1}^Kd_q^K)^{\frac{1}{2}(-1)^q}
\label{w4}
\end{eqnarray}
is the R--torsion of $d^K$ \cite{RS}. 
The R--torsion is a combinatorial invariant of $S^3\,$, i.e. it is the 
same for all choices of lattice $K$ for $S^3$ (including non-cubic, e.g.
tetrahedral, lattices). Thus when compactifying the 
spacetime ${\bf R}^3$ to $S^3$ the resulting expression (\ref{w3}) for
the partition function in the discrete theory is independent of how the 
lattice $K$ for ${\bf R}^3$ is modified to obtain a lattice decomposition
of $S^3\,$, except for the exponent of $\lambda/\pi$ in (\ref{w3}).
A straightforward calculation using the tetrahedral lattice for $S^3$
obtained by identifying $S^3$ with the boundary of the standard 4-simplex in 
${\bf R}^4$ gives $\tau(S^3;K,d^K)\!=\!1$. Thus $\widetilde{Z}_K(\lambda)\!=\!
(\pi/\lambda)^{1-N_0^K+N_1^K}$ in the present case. As in the continuum case,
if $S^3$ is replaced by a general closed oriented 3-manifold $M$ in the 
preceding then the derivation of (\ref{w3}) (with $S^3$ replaced by $M$)
continues to hold if $H^1(M)\!=\!0$ and can be generalised  
if $H^1(M)\ne0$ \cite{me}.
A deep mathematical result, proved independently
by J.~Cheeger and W.~M\"uller \cite{torsion}, states that R--torsion and
Ray--Singer torsion are equal; in particular $\tau(M;K,d^K)\!=\!
\tau_{RS}(M,d)$ (so $\tau_{RS}(S^3,d)\!=\!1$ as mentioned earlier).
It follows that the partition function 
$\widetilde{Z}_K(\lambda)$ of the discrete theory reproduces the  
continuum partition function $\widetilde{Z}(\lambda)$ 
when $\lambda\!=\!\pi\,$, and also when 
$\lambda\ne\pi$ after a lattice-dependent renormalisation of $\lambda$ in the
discrete theory.

The field doubling required in the preceding is reminiscent of the doubling
required in Thermo Field Dynamics in order that the vacuum expectation value
of an operator reproduces the statistical average \cite{thermo}.

The results of this paper are of interest in connection with lattice fermion
doubling.
From (\ref{4}) and (\ref{5}) we see that the lagrangians ${\cal L}_{CS}$
and ${\cal L}_{BF}$ of the abelian CS and BF theories can be written 
in an analogous way to the lagrangian $\psi^*\gamma^{\mu}\partial_{\mu}\psi$
for a free spinor field:
\begin{eqnarray}
{\cal L}_{CS}=A^{\dagger}e^{\mu}\partial_{\mu}A\quad,\quad
{\cal L}_{BF}=\widetilde{A}^{\dagger}\widetilde{e}^{\mu}
\partial_{\mu}\widetilde{A}
\label{21}
\end{eqnarray}
where $A\!=\!(A_{\mu})$ and $\widetilde{A}\!=\!(A,A')$ are considered as 
a 3-vector and 6-vector, $A^{\dagger}$ and $\widetilde{A}^{\dagger}$ are their
transposes, $e^{\mu}$ is a $3\times3$ matrix 
($(e^{\mu})_{\nu\rho}\!=\!\epsilon_{\nu\mu\rho}$) and $\widetilde{e}^{\mu}$
is a $6\times6$ matrix.
If we formulate the abelian CS and BF theories
on a spacetime lattice in the same way as 
for a spinor field theory and calculate the momentum space propagator in the 
standard way we find 
a ``doubling'' of exactly the same kind as for spinor fields 
on the lattice (described, e.g., in ch. 5 of \cite{Creutz(book)}).
Thus, one the one hand, when the abelian CS or BF theory is put on the lattice
in the same way as a spinor theory an analogue of ``fermion doubling'' appears,
while on the other hand the discretisation of the abelian BF theory described
in this paper successfully reproduces continuum quantities. 

The doubled, i.e. BF, version of the abelian CS theory
has the following analogue of chiral invariance.
The $6\times6$ matrix $\widetilde{C}$ defined by
$\widetilde{C}(A,A')\!=\!(A,-A')$ satisfies the chirality conditions 
$\widetilde{C}^2\!=\!I$ and $\widetilde{C}\widetilde{e}^{\mu}
\!=\!-\widetilde{e}^{\mu}\widetilde{C}$. 
Thus $\widetilde{C}$ is analogous to $\gamma^5$ in spinor 
theory, and the abelian BF lagrangian in (\ref{21}) has chiral invariance under
$\widetilde{A}\to{}e^{\alpha\widetilde{C}}\widetilde{A}$ 
($\alpha\!\in\!{\bf R}$). The original field $A$ and 
new field $A'$ then have positive- and negative chirality respectively,
in analogy with spinors of positive- and negative chirality.
(This is reminiscent of the connection 
between doubling and chirality discussed in Ref. \cite{Creutz(chirality)}.)
These observations, together with the results of this paper, suggest 
that when formulating lattice spinor theories
(in particular in the K\"ahler--Dirac framework \cite{Joos}) 
the spinors of positive- and negative chirality should
be associated with the lattice and its dual lattice respectively. 

I am grateful to Siddhartha Sen for many helpful 
discussions and encouragement. I also thank Jim Sexton and Alfredo Iorio
for helpful suggestions. This work was partially supported by FORBAIRT
and Hitachi Labs, Dublin.

\end{document}